\documentstyle[aps,twocolumn,prl,epsfig]{revtex}
\begin{document}
\title{Vortex dynamics and upper critical fields in ultrathin Bi films}\author{G. Sambandamurthy, K. Das Gupta, N. Chandrasekhar}
\address{Department of Physics, Indian Institute of Science, Bangalore 560 012, India}
\maketitle
\begin{abstract}
Current-voltage (I-V) characteristics of quench condensed, superconducting, ultrathin $Bi$ films in a magnetic field are reported. These I-V's show hysteresis for all films, grown both with and without thin $Ge$ underlayers. Films on Ge underlayers, close to superconductor-insulator transition (SIT), show a peak in the critical current, indicating a structural transformation of the vortex solid (VS). These underlayers, used to make the films more homogeneous, are found to be more effective in pinning the vortices. The upper critical fields (B$_{c2}$) of these films are determined from the resistive transitions in perpendicular magnetic field. The temperature dependence of the upper critical field is found to differ significantly from Ginzburg-Landau theory, after modifications for disorder.
\end{abstract}

\vspace{0.5cm}
\section{Introduction}
Transport properties of disordered ultrathin films have been studied extensively over the last decade especially in the context of disorder driven and magnetic field driven superconductor-insulator transition (SIT) ~\cite{allsit}. This transition is found to occur at a particular value of disorder corresponding to a critical resistance R$_c$, which clusters close to $\sim$ h/4e$^2$(the quantum of resistance for Cooper pairs), for both field driven and disorder driven transitions. Earlier studies on $Bi$, quench-condensed on underlayers of $Ge$, showed a ``homogenous" type of SIT which was characterized by a strong suppression of the critical temperature, T$_c$ as disorder was increased. The presence of a thin ($\le$ 10 $\rm \AA$) underlayer of $Ge$, is conventionally thought to improve the wetting properties of the film, and thereby assist homogeneous growth~\cite{strong}. However, several studies have indicated that the underlayer is not inert, as conventionally assumed, and may actually play an active role in determining the transport properties of the film~\cite{gerole}. Though most studies have concentrated on the transition itself, little effort has been made to understand the physics of superconductivity, such as vortex dynamics, in ultrathin films with different underlayers, which differs significantly from its manifestation in bulk material. In this work, we demonstrate some of the differences. In a previous publication, we have investiagated several underlayers for these films including solid Xenon~\cite{kdg} to investigate screening effects.

In this paper we present observations of some anomalies in the I-V's in a magnetic field and the $B_{c2}(T)$ of $Bi$ films, which are in the vicinity of the disorder driven SIT. This work compares films without and with a Ge underlayer. From current-voltage (I-V) characteristics in the superconducting state, measured in finite magnetic field, it is possible to infer the dynamics of vortices. 

\section{Experimental Results}

Experiments were carried out on films near the disorder driven SIT, with and without $Ge$ underlayers {\it simultaneously} to determine the effect of underlayers on the vortex motion in these films. Upper critical fields $B_{c2}(T)$ of these films were also measured, from the resistive transitions in perpendicular magnetic field, and their temperature dependance was found to differ significantly from Ginzburg-Landau theory~\cite{gltheory}. The experiments were done in a UHV cryostat, custom designed for {\it in-situ} experiments, described in~\cite{gsmssc}. The cryostat is pumped by a turbomolecular pump backed by an oil-free diaphragm pump. A completely hydrocarbon free vacuum $\le$ 10$^{-8}$ Torr can be attained. The substrates are amorphous quartz of size 2.5cm X 2.5cm and mounted on a copper cold finger whose temperature can be varied down to 1.8K by pumping on the liquid helium bath. The material ($Bi$) is evaporated from a Knudsen-cell with a pyrolytic Boron Nitride crucible, of the type used in Molecular Beam Epitaxy (MBE). $Bi$ is evaporated from the cell at 650$^{o}$ C, into a 4-probe resistivity measurement pattern defined by a metal mask in front of the substrate. Successive liquid helium and liquid nitrogen cooled jackets surrounding the substrate reduce the heat load on the substrate, and provide cryo-pumping. This produces an ultimate pressure $\sim$ 10$^{-10}$ Torr in the system. The flux reaching the substrate is controlled using a carefully aligned mechanical shutter in the nitrogen shield. The thickness of the film is increased by small amounts, by opening the shutter for a time interval corresponding to the desired increase in thickness. A quartz crystal thickness monitor measures the nominal thickness of the film. Electrical contacts to the film are provided through pre-deposited platinum contact pads ($\sim$ 50$\rm \AA$ thick). $Ge$ underlayers are deposited over a region of the substrate(a-quartz) before loading the substrate into the cryostat. Separate electrical connections to films on $Ge$ and on bare a-quartz allow us to study both the films simultaneously. I-V's and electrical resistance measurements are done using a standard d.c.current source (Keithley model 220/224) and nanovoltmeter (Keithley model 182) and elctrometer (Keithley model 6514). 

In this work, we have measured the critical current I$_c$ for different films at T = 2.0K. We have evaluated the power dissipated by the current in the superconducting films to be a few hundreds of nanowatts at the highest currents used (i.e. near the critical current). The cryostat has been designed to have a cooling power of at least 1 mW at this temperature. The substrates on which the films are deposited, are mounted on a copper coldfinger with silver loaded Apiezon grease, ensuring a low thermal resistance. Under such circumstances, we believe the differences in the cooling power and ohmic heating to be large enough to ensure that the temperature fluctuations due to ohmic dissipation can be safely ignored. 

When the normal state sheet resistance, R$_{\Box}^N$ of the film is close to h/4e$^2$, the combined effect of quantum fluctuations and disorder prevents the formation of a vortex solid (VS) phase, and such an absence of a VS phase has recently been reported in $Bi$ films close to SIT~\cite{valles1}. However, we find that this type of behavior can be modified when the underlayer is changed. The purpose of this paper is to present our results which establish the influence of the underlayer on the vortex phase, and others, which go beyond those reported in ref. [6]. Specifically, these are the observation of a structural phase transition in the vortex phase, which manifests itself as a `peak effect', and the significant deviations in the critical fields of these ultrathin superconducting films, when compared to Ginzburg-Landau theory.  Keeping this in mind, we study I-V characteristics in magnetic field to understand the effect of magnetic field on the vortices and how it differs for films on different substrates. In the following, after suitable preliminaries, we present our results and discuss their implications.

The I-V curves are hysteretic as shown in Fig. 1. i.e. when the current is increased from the zero value, the resistance jumps to the normal state value at a critical current, I$_c$. Upon reducing the current from the normal state, the voltage returns to zero not at I$_c$, but at a lower value. This behavior is reminescent of a underdamped Josephson junction, where the intrinsic junction capacitance, and the finite leakage resistance to quasiparticles can be modeled as shunts to an ideal junction.  The lower current in this case is the retrapping current. This suggests that our films can be considered as an array of Josephson junctions, which are shunted by an equivalent resistance and the equivalent intrinsic junction capacitance. Consequently the resistively and capacitively shunted junction (RCSJ) model~\cite{baro} has be used to describe the hysteretic behavior~\cite{gsmssc} in an earlier publication. 

Fig. 2 shows the I-V characteristics for a 26 $\rm \AA$ $Bi$ film on (a) on bare a-quartz substrates and (b) 10 $\rm \AA$ $Ge$ underlayer, both on a linear scale, at low voltages. In this figure only the positive current sweep is shown for clarity, though all the curves show hysteretic behavior. All the I-Vs are recorded at T = 2.0K. The differences are described below. In (a), at zero field, one can clearly see the dissipationless supercurrent region. Upon increasing the field, the system first becomes dissipative due to depinning of the vortices, caused by the Lorentz force acting on them due to the current I$_{dep}$, although not yet normal. At this point the I-V becomes nonlinear, and the departure from linearity is clearly seen in this figure. We note here that the definition of the depinning current is consistent with the one conventionally used in the literature~\cite{chen}. For a film of identical thickness, the I$_c$ is higher on the Be underlayer, as can be seen from Fig. 2(b). We also note that the I-V is non-linear in the vicinity of the I$_c$. The non-linearity attributable to depinning seems to set in at lower currents for the films on Ge. However, the I$_c$ for films on Ge is higher. In both these films, the dissipation is  finite at any applied field, as evidenced by a linear regime of the I-V, from which a resistance can be discerned. This is possibly due to the relatively high temperature of 2 K, when compared with published literature~\cite{chen}. This we attribute to thermally 
activated vortex motion.

Fig. 3 (a) and (b) show the I-V's on a semi-log scale, until the normal state resistance is reached, where all the curves at different magnetic fields meet.  For the films on
bare substrates, only one critical current can be unambiguously identified. For the films on Ge underlayers, two different step like increases in the voltage can be clearly identifed  for three different field values. It is only after the second step that the film resistance becomes the normal state resistance, and therefore this should be defined as the critical current. Hence the first step has to be associated with some sort of structural transition in the vortex solid. Recall that the depinning current is lower than this current at which the first step in the voltage appears.

Fig. 4 shows the critical current as a function of normalized temperature, at zero field for films on Ge as well as on bare quartz.  A peak in the critical current in the vicinity of the transition temperature is clearly visible, for the films on Ge. 
An identical phenomenon has been extensively invesitaged in layered superconductors such as 2H-NbSe$_2$~\cite{peak1} and several other materials such as amorphous Nb$_3$Ge and Mo$_3$Si films.~\cite{peakes} This has been given the name "peak effect". The pinning force density F$_p$ exhibits a prominent peak slightly below the upper critical field, B$_{c2}$ or near the critical temperature. This effect is attributed to elastic instabilities generated by local fluctuations of the pinning forces, which induce a rapid softening of the flux line solid.  This softening allows the flux lines to conform readily to a configuration that locks it to the inhomogeneities, so that the critical current I$_c$ increases ~\cite{peak2}. So the observation of a peak in I$_c$ implies that an elastic flux line solid transforms to a plastic solid in the vicinity of the peak. For the films on Ge, we identify the first transition with the "peak effect".  On the other hand, for films on bare substrate at B = 0 G, the system shows the usual depinning transition with some finite dissipation and non-linear I-V's, until I$_c$, where it becomes normal. In our films, the critical fields are higher than what we can attain using our magnet. Hence we have been unable to observe the peak effect in the I-H plane. However, the observation in the I-T plane is unambiguous proof of the peak effect~\cite{shobho}.

There is another important difference between the films on $Ge$ and bare substrates. The I$_c$ when plotted against the applied field B, shows a maximum for films on bare substrates as shown in Fig. 5. This effect is therefore real, as the peak is much larger than the error bars. Such an effect has been observed in Josephson junction arrays, where the cirtical current shows a dependence on the field for small magnetic fields~\cite{chen}. In such JJ array structures, a relevant variable is the number of free vortices per unit cell, and is denoted as f ($f=B{a_o}^2/\Phi_o$), where B is the applied field. Since we have sufficient reason to believe that our films
can be understood as a random JJ array as discussed above~\cite{gsmssc,kdg,gsm2}, this reasonable explanation for this effect. From the measured I$_c$, and using the Ambegaokar-Baratoff relation~\cite{ab}, we can estimate the number of junctions in our film
to be a million. Knowledge of the applied field, and using our published RHEED data~\cite{gsmssc,kdg,gsm2} to estimate a$_o$ (of the order of 2000 $\AA$), gives us a value for f of 0.4, which is consistent with the values reported in the literature. It is intriguing that such a frustration effect is not observed for films on Ge underlayers. We do not know the reason for this at this time. We are however in the process of investigating other films and underlayers in order to aid our understanding of this phenomenon.

We now discuss the critical fields of these films. The temperature dependence of the upper critical field, B$_{c2}$ of a typical type II superconductor can be shown, by Ginzburg-Landau theory to be of the form ~\cite{tink},
\begin{equation}B_{c2} = B_{c2}(0)(1-t^2)
\end{equation}
where B$_{c2}$(0) is the upper critical field at T=0 and t 
(= T/T$_c$) is the reduced temperature. Close to the transition temperature (t $\sim$ 1), the relation above can be approximated by $(1-t)$.

Fig.6 shows a representative B$_{c2}$ vs $t$ plot for a 26 $\rm \AA$ film where the data points are indicated by triangles, and the solid line is a fit discussed below. The data show a curvature opposite to that expected by GL theory near T$_c$. The data is found to fit the following equation,\begin{equation}B_{c2} = B_{c2}(0)(1-t)^{\alpha}\end{equation}where $\alpha$ = 1.175. Even at $t\simeq 1$, we do not observe the predicted linear dependence. In Fig.6, the solid line is a fit to equation 3 with $\alpha$ = 1.175, the long-dashed line is the GL expression, medium-dashed line is the linear dependence, and the short-dashed line is  $\alpha$ = 1.45. This last value of $\alpha$ is found to characterize the melting line in extreme type II superconductors, such as the high T$_c$ materials~\cite{melt}. All the films studied (thickness ranging from 26 to 105 $\rm \AA$) show the same curvature with $\alpha$ $\sim$ 1.14 $\pm$ 0.05.

\section{Discussion}

The transport properties in films such as the ones studied here are examples of transport in heterogeneous media~\cite{stauffer,orbach,prester}, i.e. as the problems of dynamics of percolation networks. The current path in such films is clearly percolative, and one cannot rule out such effects in the motion of vortices. A possible model for these films seems to
be a random composite network (RCN), in which both superconducting and normal regions exist~\cite{efros}. Attempts have been made to quantitatively interpret the I-V's in such
structures~\cite{prester}. Currently we are in the process of incorporating the two dimensional nature of our system, and the effects of a magnetic field to quantitatively understand our I-V's. 

The I-V's are sensitive to the deposition temperature. All the I-V's discussed above were obtained from films for which the temperature of deposition was 15 K. Several dramatic 
differences are seen in the I-V's of films that are deposited at 4 K or 1.5 K. These results along with a quantitative model for the I-V's will be addressed in a separate publication.

For a disordered superconductor, the temperature dependence of the critical field gets modified due to the interplay between pinning, and thermal fluctuations. In a weakly disordered superconductor, the temperature dependence of the upper critical field shows significant deviations from this relation only at low temperatures. The interplay between pinning and fluctuations increases the critical fields. Such increases have been successfully explained by the Werthamer-Helfand-Hohenberg (WHH) theory~\cite{wwh}, which is based on the Ginzburg-Landau formalism.However, the WHH model has not been successful in explaining the B$_{c2}(T)$ of amorphous transition metal alloys~\cite{tran} which often show an upward curvature. 

Later models taking into consideration, the effect of localization and electron-electron interactions~\cite{mef}, were successful in explaining the upward curvature of the B$_{c2}$ at lower temperatures. We have determined B$_{c2}$(T) of our films, from the resistive transitions in a perpendicular magnetic field. The convention that we have followed is to define B$_{c2}$ as the field at which the sheet resistance is half its normal state value, R$_{\Box}^N$, which is the generally accepted convention. The current values used to record the resistive transition of the films are small ($\sim$ few nano amperes) and the magnetic field produced by them is extremely small (milligauss) compared to the critical fields (Tesla) of these films. 

Although the theories discussed above~\cite{wwh,mef} explain the upward curvature of B$_{c2}$ at lower temperatures very well, they predict only a linear dependence (i.e. $(1-t)$) of B$_{c2}$ near T$_c$. None of  them has any explanation for this peculiar behavior observed in our films. A power of $\sim$ 1.14, which is between the value of 1.45 for a melting line and 1.0 for GL theory suggests that a new explanation may be required to understand this data.  The Pauli limiting field or the Clogston-Chandrasekhar limit for disordered superconductors is given by H$_p$ = 1.84 T$_c$ Tesla~\cite{pauli}. The value of H$_p$ is approximately 5.3 T even the for the thinnest films studied here. In our experiments we have not crossed that field at any time. 

\section{Conclusions}

In conclusion, the I-V's of quench condensed Bi films with and without underlayers of $Ge$ show evidence for very interesting vortex dynamics. In particular, we
have presented data that show a peak effect, which has been observed for the first time
in such ultrathin films. We also show effects that can be attributed to frustration in a
random JJ array. These effects are sensitive to the microstructure of the film, and the
underlayers used to deposit the film. The upper critical fields are found to have a power law temperature dependence with a power of $\sim$ 1.14, which needs further investigation.  This, together with the peak effect and frustration effects, suggests a novel behavior in such ultrathin disordered films, which has not been addressed.
More experiments on highly disordered films with R$_{\Box}^N$ $\sim$ h/4e$^2$ are needed to completely understand the vortex dynamics. 

This work is supported by Department of Science and Technology, and UGC Government of India. One of the authors (KDG) thanks the Council of Scientific and Industrial Research, New Delhi for the fellowship. NC thanks Dr. A. M. Goldman and Dr. S. Bhattacharya for discussions.

\newpage{\centerline {FIGURE CAPTIONS}}

(1) Fig. 1. I-V characteristics of a 26 $\rm \AA$ film on $Ge$ underlayer showing hysteresis for (1) zero field (2) 364 G and (3) 7981G. The curves were recorded at T = 2.0K.\\

(2) Fig. 2. I-V characteristics of 26 $\rm \AA$ $Bi$ film on (a) bare quartz and (b) on a 10 $\AA$ $Ge$ underlayer on a linear scale at low voltages. All I-Vs were recorded at T = 2.0K. Different curves indicate different field values. \\

(3) Fig. 3. I-V characteristics of 26 $\rm \AA$ $Bi$ film on (a) bare quartz and (b) on a 10 $\AA$ $Ge$ underlayer on a semi-log scale, taken until the films are driven normal by te current. All I-Vs were recorded at T = 2.0K. Different curves indicate different field values. \\

(4) Fig. 4. Temperature dependance of the critical current (I$_c$) for films on bare substrates and films on Ge underlayers. \\

(5) Fig. 5. Field dependance of the critical current (I$_c$) for the film of Fig. 2 (b). \\

(6) Fig. 6. B$_{c2}$ vs $t$ for a 26 $\rm \AA$ film. The solid line is the fit to equation 3 with $\alpha$ = 1.175. See text for discussion.\\

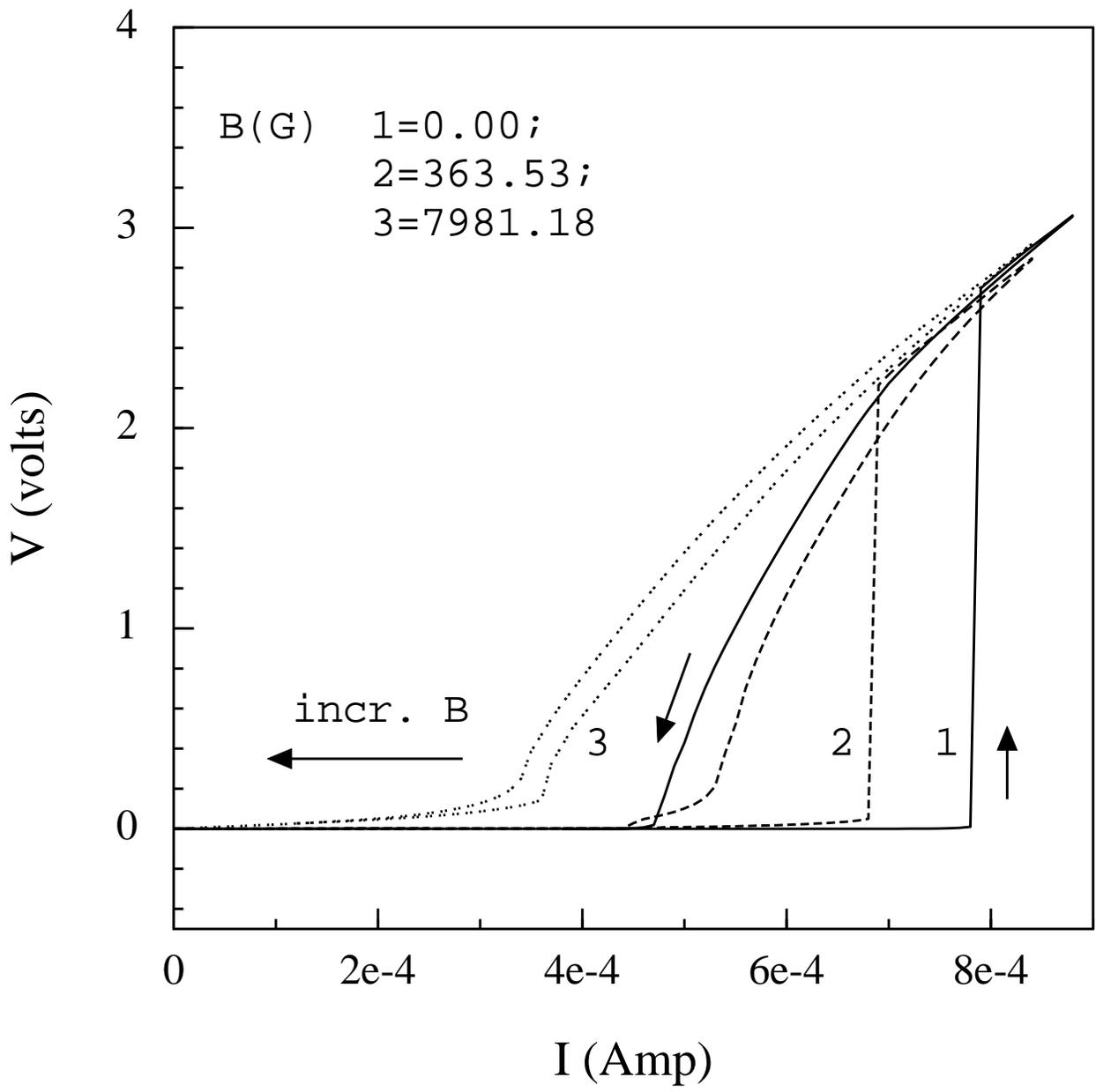
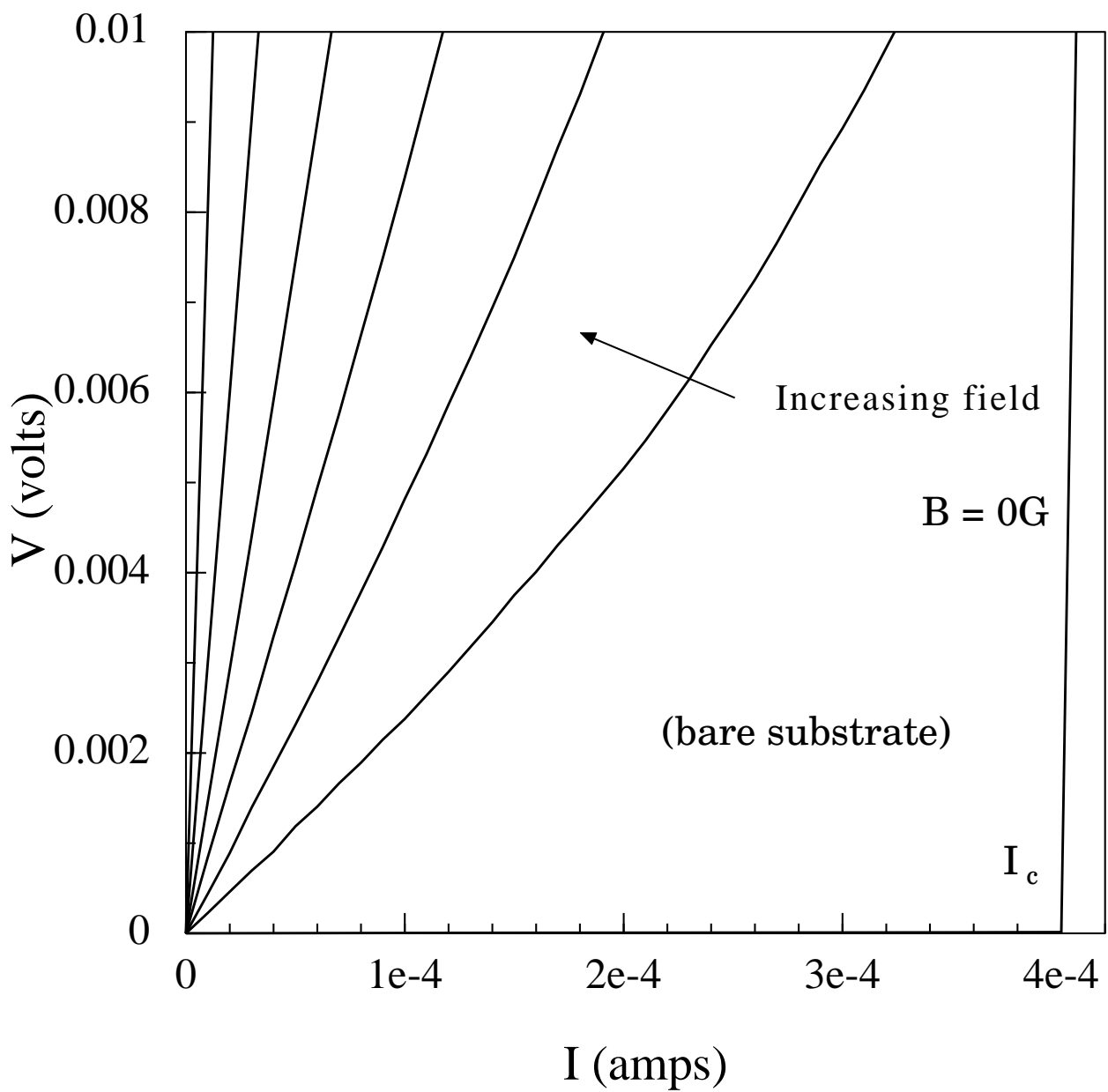
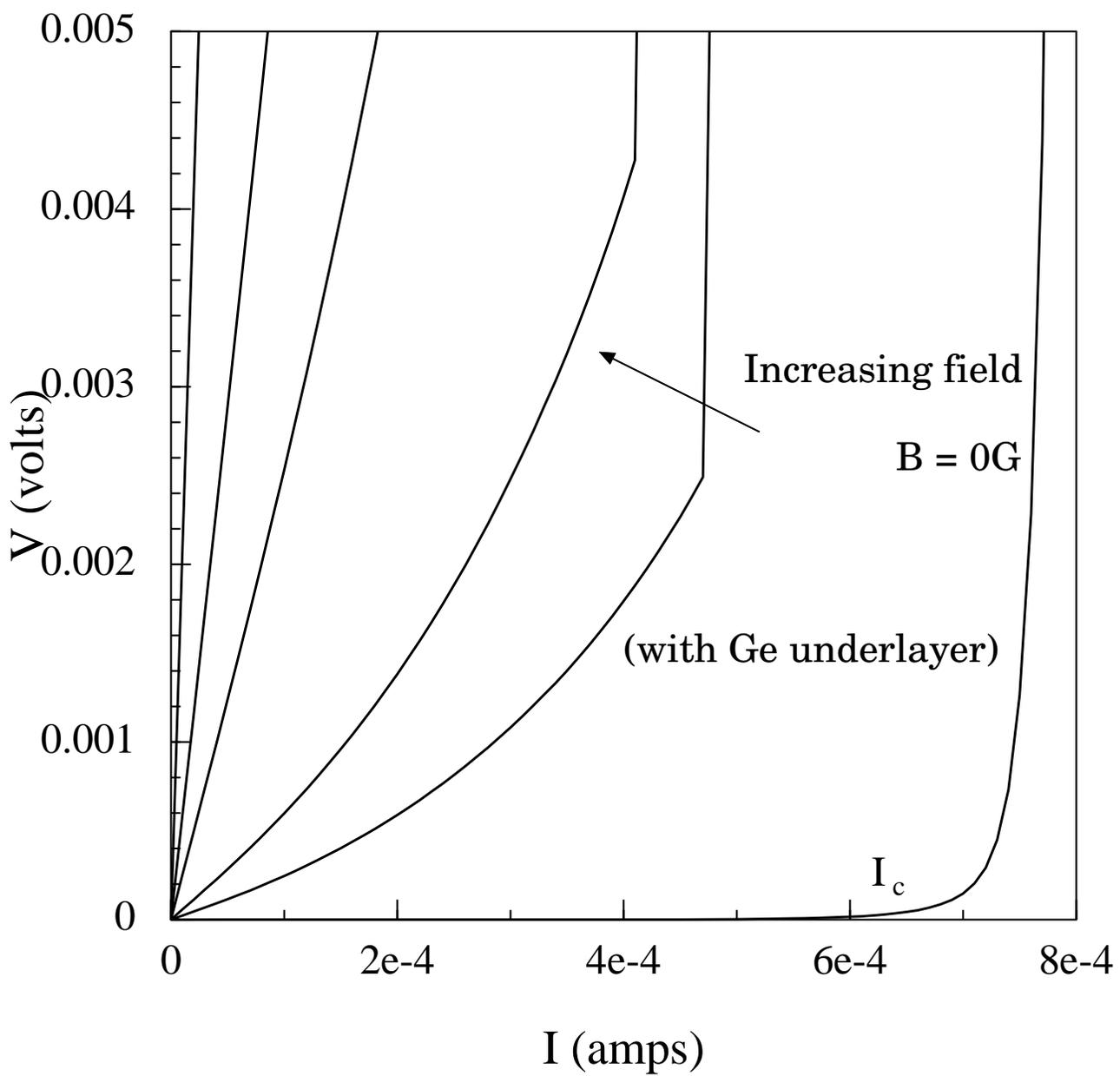
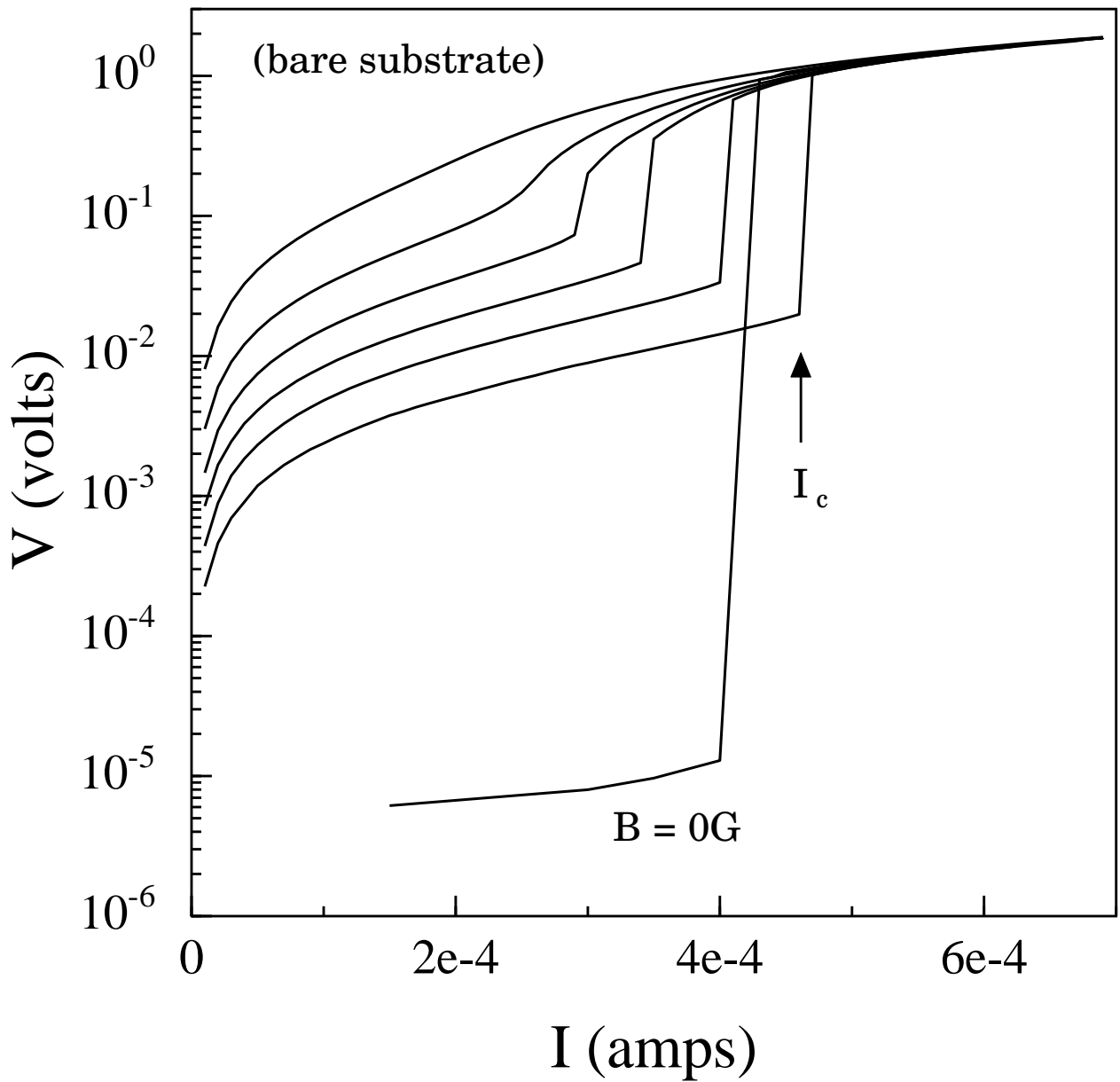
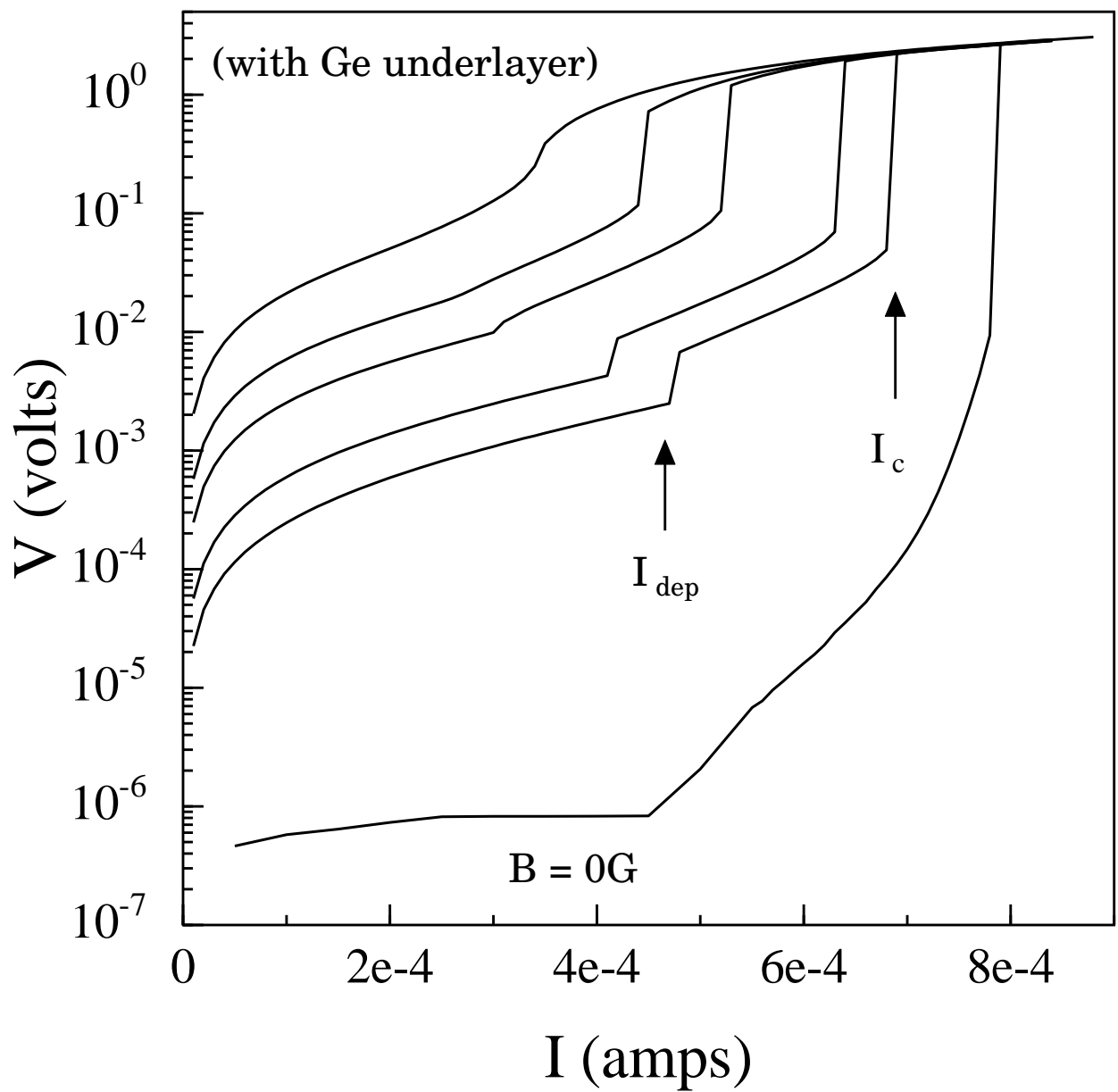
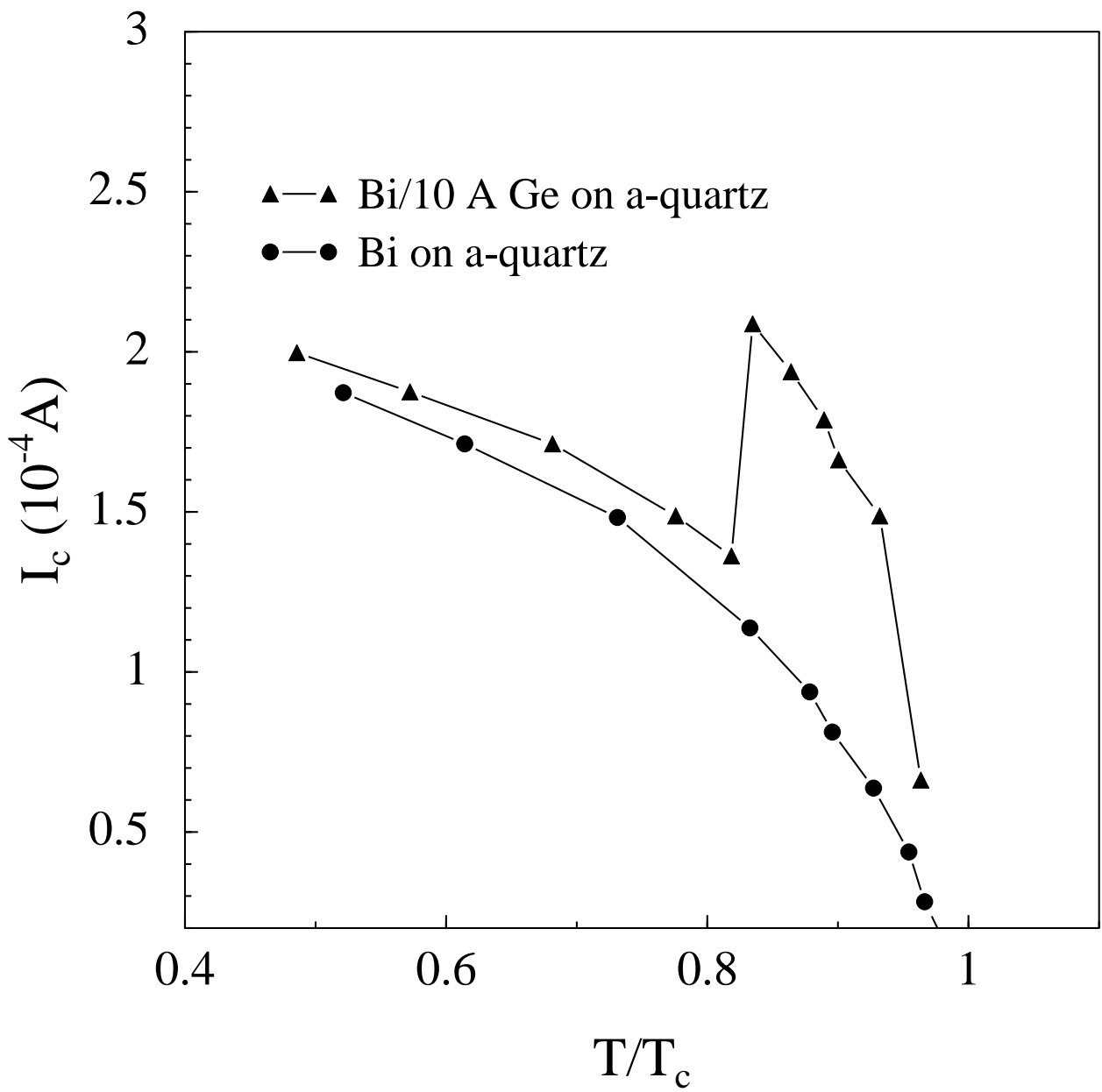
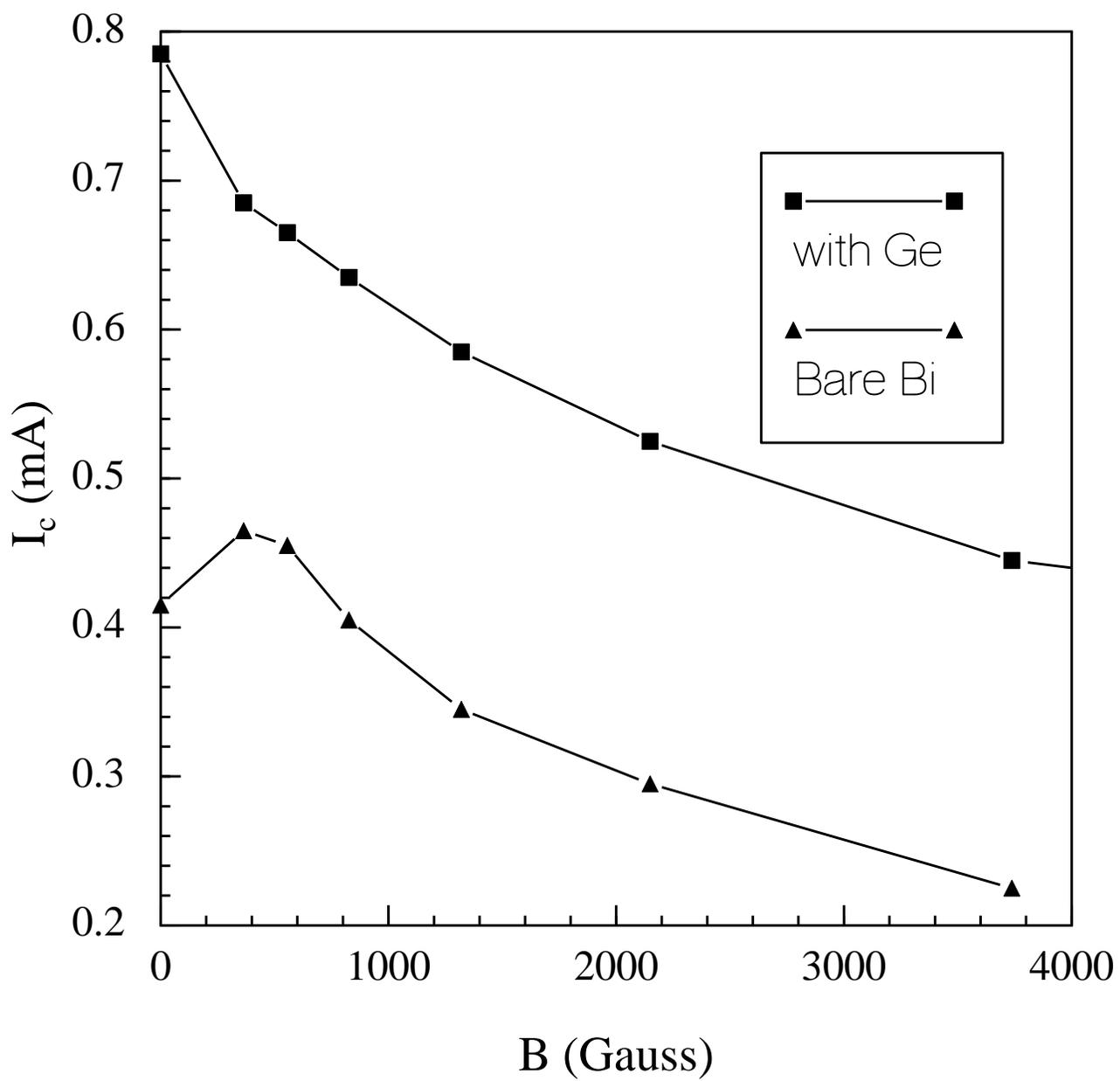
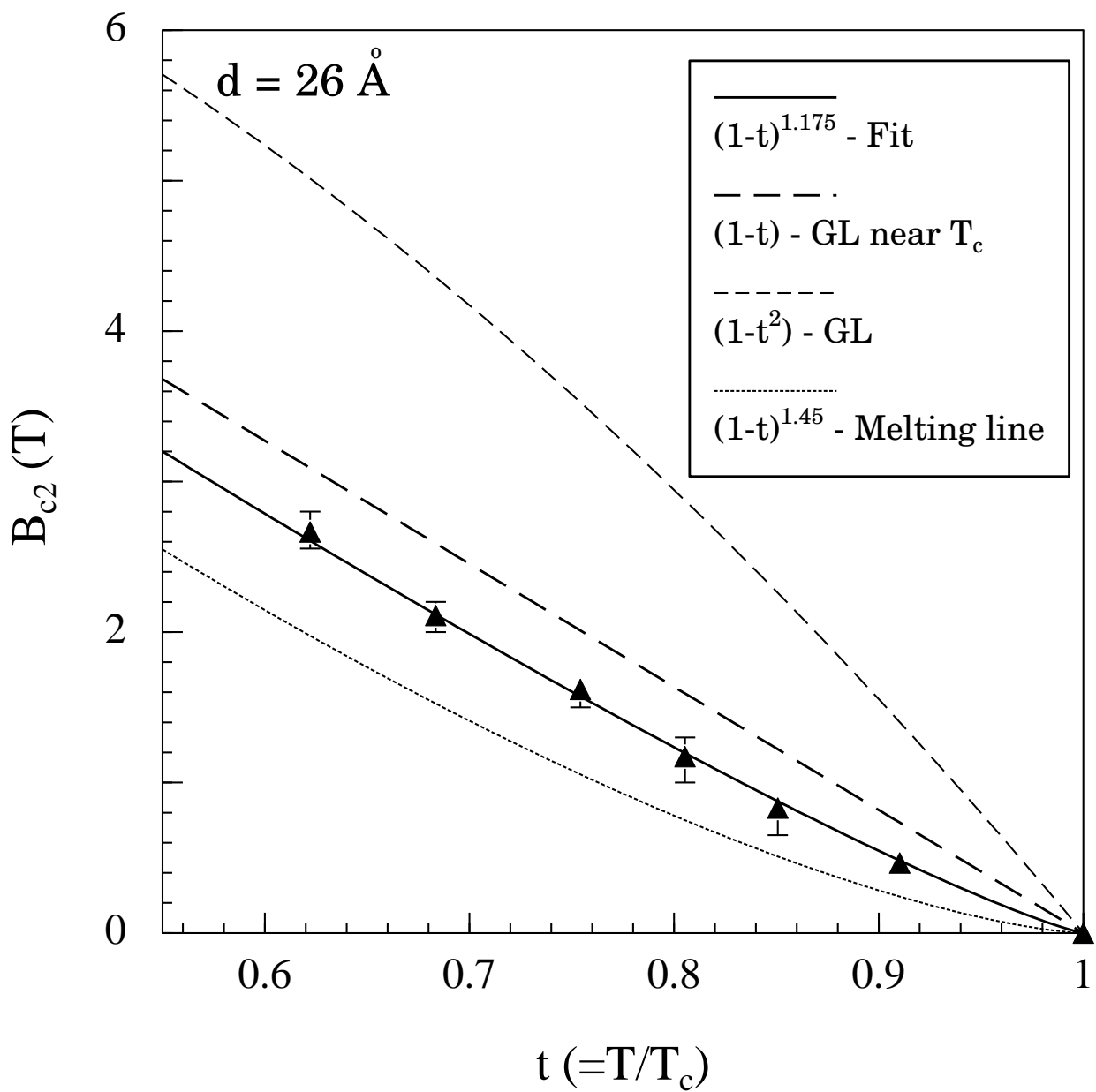
\end{document}